\documentclass[graybox,vecmaths]{svmult}

\usepackage{mathptmx}       
\usepackage{helvet}         
\usepackage{courier}        
\usepackage{type1cm}        
\usepackage{makeidx}         
\usepackage{graphicx}        
\usepackage{multicol}        
\usepackage[bottom]{footmisc}

\makeindex             

\def\espaitemps{({\cal V},g)}

\def\be{\begin{equation}}
\def\ee{\end{equation}}
\def\bea{\begin{eqnarray}}
\def\eea{\end{eqnarray}}
\def\bean{\begin{eqnarray*}}
\def\eean{\end{eqnarray*}}

\begin{document}

\title*{Remarks on the stability operator for MOTS}
\titlerunning{Stability operator for MOTS}
\author{Jos\'e M. M. Senovilla}
\institute{J.M.M. Senovilla \at F\'{\i}sica Te\'orica, Universidad del Pa\'{\i}s Vasco, Apartado 644, 48080 Bilbao, Spain,
\email{josemm.senovilla@ehu.es}}
%
%
\maketitle

\abstract*{Small deformations of marginally outer trapped surfaces (MOTS) are studied by using the stability operator introduced by 
Andersson-Mars-Simon. Novel formulae for the principal eigenvalue 
are presented. A characterization of the many marginally outer 
trapped tubes passing through a given MOTS are given, and 
the possibility of selecting a privileged MOTT is 
discussed. This is related to the concept of `core' of a black hole: a 
minimal region that one should remove from the spacetime in order to 
get rid of all possible closed trapped surfaces. In spherical symmetry 
one can prove that the spherical MOTT is the boundary of a core. I 
argue how similar results may hold in general black-hole spacetimes.}


\section{Basic concepts and the stability operator}
\label{sec:1}

Let $S$ denote a closed {\em marginally outer trapped surface} (MOTS) in the spacetime $\espaitemps$, so that its outer null expansion vanishes $\theta_{\vec k}=0$ \cite{HE,Wald}. 
Here, the two future-pointing null vector fields orthogonal to $S$ are denoted by $\vec{\ell}$ and $\vec k$ and we set $\ell^\mu k_{\mu}=-1$.
I will also use the concept of OTS ($\theta_{\vec k} <0$ ). A {\it marginally (outer) trapped tube} (MOTT) is a hypersurface foliated by MOTS.

As proven in \cite{AMS1}, the variation $\delta_{f\vec n} \theta_{\vec k}$ of the vanishing expansion along any normal direction $f\vec n$ such that $k_\mu n^\mu=1$ reads
\be
\delta_{f\vec n} \theta_{\vec k}=-\Delta_{S}f+2s^B\overline\nabla_{B}f+
f\left(K_{S}-s^B s_{B}+\overline\nabla_{B}s^B-\left.G_{\mu\nu}k^\mu \ell^{\nu}\right|_S -\frac{n^\rho n_{\rho}}{2}\,  W\right)
\label{deltatheta}
\ee
where $K_{S}$ is the Gaussian curvature on $S$, $\Delta_{S}$ its Laplacian, $G_{\mu\nu}$ the Einstein tensor, $\overline\nabla$ the covariant derivative on $S$, $s_{B}=k_{\mu}e^\sigma_{B}\nabla_{\sigma}\ell^\rho$ (with $\vec e_{B}$ the tangent vector fields on $S$), and $W\equiv \left.G_{\mu\nu}k^\mu k^{\nu}\right|_S +\sigma^2$
with $\sigma^2$ the shear scalar of $\vec k$ at $S$. 
Note that the direction $\vec n$ is selected by fixing its norm $
\vec n =-\vec\ell +\frac{n_{\mu}n^{\mu}}{2}\vec k$
and that the causal character of $\vec n$ is unrestricted.
Under usual energy conditions \cite{HE,Wald} $W\geq 0$ and actually $W=0$ can only happen if $\left.G_{\mu\nu}k^\mu k^{\nu}\right|_S=\sigma^2=0$ leading to Isolated Horizons \cite{AK1}, so that I shall assume $W>0$ throughout.

The righthand side in (\ref{deltatheta}) defines a linear differential operator $L_{\vec n}$ acting on $f$: $\delta_{f\vec n} \theta_{\vec k}\equiv L_{\vec n} f$.
$L_{\vec n}$ is an elliptic operator on $S$, called \underline{the stability operator} for $S$ in the normal direction $\vec n$. $L_{\vec n}$ is not self-adjoint in general (with respect to the $L^2$-product on $S$). Nevertheless, it has a real principal eigenvalue $\lambda_{\vec n}$, and the corresponding (real) eigenfunction $\phi_{\vec n}$ can be chosen to be positive on $S$.
The (strict) stability of the MOTS $S$ along a spacelike $\vec n$ is ruled by the (positivity) non-negativity of $\lambda_{\vec n}$.

The formal adjoint operator with respect to the $L^2$-product on $S$ is given by
$$
L^\dag_{\vec n} \equiv -\Delta_{S}- 2s^B\overline\nabla_{B}+\left(K_{S}-s^B s_{B}-\overline\nabla_{B}s^B-\left.G_{\mu\nu}k^\mu \ell^{\nu}\right|_S -\frac{n^\rho n_{\rho}}{2}\,  W\right)
$$
and has the same principal eigenvalue $\lambda_{\vec n}$ as $L_{\vec n}$  \cite{AMS1}. I denote by $\phi^\dag_{\vec n}$ the corresponding principal (real and positive) eigenfunctions.

\section{Many MOTTs through a single MOTS}
For each normal vector field $\vec n$, the operator $L_{\vec n}-\lambda_{\vec n}$ has obviously a vanishing principal eigenvalue (and the same principal eigenfunction $\phi_{\vec n}$). This operator $L_{\vec n}-\lambda_{\vec n}$ corresponds to the stability operator $L_{\vec n'}$ along another normal direction $\vec n'$ given by
$n'^\mu n'_\mu =n^\mu n_\mu +(2/W) \lambda_{\vec n}$, so that $\delta_{\phi_{\vec n}\vec n'} \theta_{\vec k}=0$. If $\vec n$ is spacelike and $S$ is strictly stable along $\vec n$ ($\lambda_{\vec n} > 0$), then $\vec n'$ points ``above'' $\vec n$ (having $n'^\mu n'_\mu > n^\mu n_\mu$).
As is obvious, the directions tangent to MOTTs through $S$ are contained in the set of such primed directions $\{\phi_{\vec n}\vec n'\}$. These MOTTs will generically be different. In fact, given two arbitrary normal vector fields $\vec n_1$ and $\vec n_2$
one can easily prove that the corresponding ``primed'' directions are equal (so that the local MOTTs coincide) if, and only if, $
\vec{n}_{1}-\vec{n}_{2} =\frac{\mbox{const.}}{W} \vec k$.
On the other hand, for any two normal vector fields $\vec n_1$ and $\vec n_2$
\be
(W/2) f\left(n_1^\rho n_{1\rho}-n_2^\rho n_{2\rho}\right)=\left(L_{\vec n_2}-L_{\vec n_1} \right)f
\label{n1n2}
\ee
providing the relation between two deformation directions pointwise.

For any given $\vec n$ one easily gets
$$
\oint_S L_{\vec n} f=\oint_S \left(K_{S}-s^B s_{B}-\overline\nabla_{B}s^B-\left.G_{\mu\nu}k^\mu \ell^{\nu}\right|_S -\frac{n^\rho n_{\rho}}{2}\,  W\right)f
$$
$$
\oint_S L^\dag_{\vec n} f=\oint_S \left(K_{S}-s^B s_{B}+\overline\nabla_{B}s^B-\left.G_{\mu\nu}k^\mu \ell^{\nu}\right|_S -\frac{n^\rho n_{\rho}}{2}\,  W\right)f
$$
in particular for the principal eigenfunctions
$$
\lambda_{\vec n} \oint_S \phi_{\vec n} =\oint_S  \left(K_{S}-s^B s_{B}-\overline\nabla_{B}s^B-\left.G_{\mu\nu}k^\mu \ell^{\nu}\right|_S -\frac{n^\rho n_{\rho}}{2}\,  W\right) \phi_{\vec n}
$$
$$
\lambda_{\vec n} \oint_S \phi^\dag_{\vec n} =\oint_S  \left(K_{S}-s^B s_{B}+\overline\nabla_{B}s^B-\left.G_{\mu\nu}k^\mu \ell^{\nu}\right|_S -\frac{n^\rho n_{\rho}}{2}\,  W\right) \phi^\dag_{\vec n}
$$
which are two explicit {\em formulas} for the principal eigenvalue
{\em bounding} it
\bea
\min_{S} \left(K_{S}-s^B s_{B}\pm\overline\nabla_{B}s^B-\left.G_{\mu\nu}k^\mu \ell^{\nu}\right|_S -\frac{n^\rho n_{\rho}}{2}\,  W\right) \leq \lambda_{\vec n} \nonumber\\
\leq \max_S  \left(K_{S}-s^B s_{B}\pm\overline\nabla_{B}s^B-\left.G_{\mu\nu}k^\mu \ell^{\nu}\right|_S -\frac{n^\rho n_{\rho}}{2}\,  W \right) \, .
\label{lambdazbounds}
\eea
Furthermore, the two functions
$\lambda_{\vec n} - \left(K_{S}-s^B s_{B}\pm\overline\nabla_{B}s^B-\left.G_{\mu\nu}k^\mu \ell^{\nu}\right|_S -\frac{n^\rho n_{\rho}}{2}\,  W\right)$ {\em must vanish} somewhere on $S$ for all $\vec n$.
 
There are two obvious simple choices $\vec n_{\pm}$ leading to a vanishing principal eigenvalue: $n^\mu_\pm n_{\pm \mu}=\frac{2}{W}\left(K_{S}-s^B s_{B}\pm\overline\nabla_{B}s^B-\left.G_{\mu\nu}k^\mu \ell^{\nu}\right|_S \right)$. 
The corresponding stability operators are $
L_+ = -\Delta_S +2s^B\overline\nabla_B$ and $L_- =  -\Delta_S +2s^B\overline\nabla_B +2\overline\nabla_B s^B$.
Denoting by $\phi_{\pm}>0$ the corresponding principal eigenfunctions one has $L_{\pm} \phi_{\pm} =0$.
The respective formal adjoints read:
$L^\dag_+ = -\Delta_S -2s^B\overline\nabla_B-2\overline\nabla_B s^B$ and $L^\dag_- =  -\Delta_S -2s^B\overline\nabla_B$
with vanishing principal eigenvalues too. Observe that $L_-$ and $L^\dag_+$ are gradients $L_- f =-\overline\nabla_{B}\left(\overline\nabla^B f-2fs^B \right)$,\, \, \, 
$L^\dag_+ f = -\overline\nabla_{B}\left(\overline\nabla^B f+2fs^B \right)$.

\section{A distinguished MOTT}
The previous property distinguishes $L_-$ as having special relevant properties, because (\ref{n1n2}) leads to
\be
\fbox{$\displaystyle{
(W/2)f\left(n^\rho n_{\rho}-n_-^\rho n_{- \rho}\right)=
L_- f -\delta_{f\vec n}\theta_{\vec k}}$} \label{n-m2}
\ee
For any other direction $\vec n'$ defining a local MOTT
$$
(W/2)\left(n'^\rho n'_{\rho}-n_-^\rho n_{- \rho}\right)=
\lambda_{\vec n} - \left(K_{S}-s^B s_{B}-\overline\nabla_{B}s^B-\left.G_{\mu\nu}k^\mu \ell^{\nu}\right|_S -\frac{n^\rho n_{\rho}}{2}\,  W\right)
$$
and, as remarked above, the righthand side must change sign on $S$.
\begin{theorem}
The local MOTT defined by the direction $\vec n_-$ is such that any other nearby local MOTT must interweave it: the vector $\vec n' -\vec n_- (\propto \vec k)$ changes its causal orientation on any of its MOTSs.
\end{theorem}

From (\ref{n-m2}), deformations using $c\phi_-$ with constant $c$  lead to outer untrapped (resp. trapped) surfaces if $c\left(n^\rho n_{\rho}-n_-^\rho n_{- \rho}\right) <0$ (resp. $> 0$) everywhere. Integrating (\ref{n-m2}) on $S$ one thus gets
$$
\frac{1}{2}\oint_{S} W f \left(n^\rho n_{\rho}-n_-^\rho n_{- \rho}\right) =-\oint_{S} \delta_{f\vec n}\theta_{\vec k}
$$
hence the deformed surface can be outer trapped (untrapped) only if $f\left(n^\rho n_{\rho}-n_-^\rho n_{- \rho}\right) $ is positive (negative) somewhere. 
If the deformed surface has $f\left(n^\rho n_{\rho}-n_-^\rho n_{- \rho}\right) < 0$ (respectively $ > 0$) everywhere then $\delta_{f\vec n}\theta_{\vec k}$ must be positive (resp.\ negative) somewhere. 

Choose the function $f=a_{0}\phi_- +\tilde f$ for a constant $a_{0}>0$ so that, as $\phi_- >0$ has vanishing eigenvalue, (\ref{n-m2}) becomes
$
(W/2)(a_{0}\phi_- +\tilde f) \left(n^\rho n_{\rho}-n_-^\rho n_{- \rho}\right)=L_-\tilde f-\delta_{f\vec n}\theta_{\vec k}$.
This can be split into two parts:
\be
(W/2)a_{0}\phi_- \left(n^\rho n_{\rho}-n_-^\rho n_{- \rho}\right)=-\delta_{f\vec n}\theta_{\vec k}, \hspace{3mm}
\frac{W}{2}\left(n^\rho n_{\rho}-n_-^\rho n_{- \rho}\right)=\frac{L_-\tilde f }{\tilde f }\label{second}
\ee
The first of these tells us that $\delta_{f\vec n}\theta_{\vec k}<0$ whenever $\vec n$ points ``above'' $\vec n_-$. But then the second in (\ref{second}) requires finding a function $\tilde f$ such that $L_-\tilde f/\tilde f$ is strictly positive on $S$.
This leads to the following interesting mathematical problem:
\begin{quotation}
Is there a function $\tilde f$ on $S$ such that (i)
$L_- \tilde f/\tilde f \geq \epsilon >0$, (ii) $\tilde f$ changes sign on $S$, and (iii) $\tilde f$ is positive in a region as small as desired?
\end{quotation}
To prove that there are OTSs penetrating both sides of the MOTT it is enough to comply with points (i) and (ii). If the operator $L_-$ has any real eigenvalue other than the vanishing principal one, then these two conditions do hold for the corresponding real eigenfunction because integration of $L_-\psi = \lambda \psi$ implies $\oint_S \psi =0$ (as $\lambda >0$) ergo $\psi$ changes sign on $S$. However, even if there are no other real eigenvalues the result might hold.
Point (iii) would ensure, then, that the deformed OTS intersects the trapped region ``above'' the MOTT only in a portion that can be shrunk as much as desired. This is important for the concept of {\em core} and its boundary, see \cite{BS1}.

As illustration of the above, consider a marginally trapped round sphere $\varsigma$ in a spherically symmetric space-time, that is, any sphere with $r=2m$ where $4\pi r^2$ is its area and $m=(r/2) (1-r_{,\mu}r^{,\mu})$ is the ``mass function''.
For any such $\varsigma$, $s^B =0$ and $\sigma^2=0$, ergo the directions $\vec n_{\pm}$ and operators $L_{\pm}$ coincide:
$\vec n_+=\vec n_- \equiv \vec m$, $L_+=L_- =L_{\vec m}= -\Delta_\varsigma$. As it happens, $\vec m$ is tangent to the unique spherically symmetric MOTT: $r=2m$ \cite{BS1}. Therefore, points (i) and (ii) are easily satisfied by choosing $\tilde f$ to be an eigenfunction of the spherical Laplacian $\Delta_{\varsigma}$, say $\tilde f =c P_{l}$ for a constant $c$ and $l>0$, where $P_{l}$ are the Legendre polynomials. Actually, one can find an explicit function satisfying point (iii) too, proving that the region $r\leq 2m$ is a core in spherical symmetry, \cite{BS1}. This is a surprising, maybe deep result, because the concept of core is global and requires full knowledge of the future, however its boundary $r=2m$ is a MOTT, hence defined locally. Whether or not this happens in general is an open important question.

{\bf Acknowdledgments:}
Supported by grants FIS2010-15492 (MICINN), GIU06/37 (UPV/EHU) and P09-FQM-
4496 (J. Andaluc\'{\i}a--FEDER) and UFI 11/55 (UPV/EHU).

\end{document}